\documentclass[aps,floatfix,preprint,preprintnumbers,nofootinbib,superscriptaddress,natbib]{revtex4}

\usepackage[titletoc]{appendix}
\usepackage{graphicx,float}
\usepackage{epsfig}		
\usepackage{dcolumn}
\usepackage{bm}
\usepackage{subfigure}
\usepackage[all]{xy}
\usepackage{amsmath,upgreek}
\usepackage{amssymb}
\usepackage{pdfpages}
\usepackage{color}
\usepackage{graphicx,epstopdf}
\usepackage[colorlinks=true,
 linkcolor=red,
 urlcolor=purple,
 citecolor=blue,hyperindex]{hyperref}

\def\0{\mbox{\tiny $0$}}
\def\1{\mbox{\tiny $1$}}
\def\2{\mbox{\tiny $2$}}
\def\3{\mbox{\tiny $3$}}
\def\4{\mbox{\tiny $4$}}
\def\5{\mbox{\tiny $5$}}
\def\6{\mbox{\tiny $6$}}
\def\7{\mbox{\tiny $7$}}
\def\8{\mbox{\tiny $8$}}
\def\9{\mbox{\tiny $9$}}

\def\f14{\mbox{\tiny $\frac{1}{4}$}}

\begin{document}

\title{Soft quantum back reaction to the Hubble tension: a smeared-out early time cosmological energy density}
\author{A. E. Bernardini}
\email{alexeb@ufscar.br}
\altaffiliation[On leave of absence from]{~Departamento de F\'{\i}sica, Universidade Federal de S\~ao Carlos, PO Box 676, 13565-905, S\~ao Carlos, SP, Brasil.}
\affiliation{Departamento de F\'isica e Astronomia, Faculdade de Ci\^{e}ncias da
Universidade do Porto, Rua do Campo Alegre 687, 4169-007, Porto, Portugal.}
\date{\today}
 
\begin{abstract}
A theoretical explanation for the so-called Hubble tension is provided within the framework of phase-space quantum mechanics extended to (quantum) cosmology. Following a description of the overall nature of this tension, with due attention to recent observational developments, a quantum cosmology framework based on the Weyl-Wigner phase-space quantum approach is presented. This circumvents the discrepancy between early- and late-time Universe predictions for the Hubble constant, $H_0$. The emergence of quantum-origin corrections dependent on a single parameter --- mediated by (generic) localized phase-space quantum states free of data analysis ---  yields predictions for $H_0$ that smoothly interpolate between early- and late-time phenomenological values, thereby joining the plethora of solutions for the Hubble tension.
\end{abstract}

\date{\today}
\maketitle

\section{Introduction}

The Hubble expansion rate, $H$, and the respective determination of the Hubble constant, $H_0$, are the cornerstone of the cosmological standard model, the so-called $\Lambda$CDM framework. $H_0$ is the ratio of the recessional velocity to distance for the observed galaxies which corresponds to the cosmic expansion rate at present, such that $H/H_0 = \dot{a}/a$, where $a\equiv a(t)$ is the time dependent scale factor. The refinement of the model parameters constrained by the experimental data analysis performed over the last decade has suggested conflicting $H_0$ values whether directly determined from measurements of distance and redshift or, instead, from the $\Lambda$CDM calibrated by measurements from the early Universe.

Therefore, such a {\em Hubble tension} problem encompasses an apparent incompatibility between local measurements of the current expansion rate of the Universe ($H_0$) when measurements from {\em early} and {\em late} times are confronted.
Initially, it could be summarized, but not restricted, by the inconsistency between the value inferred from Planck collaboration's observation of the cosmic microwave background (CMB), which is supported by the $\Lambda$CDM model for the {\em early} Universe and predicts $H_0 = (67.27 \pm 0.60) \, km/(s. Mpc)$ \cite{4444}, and the value measured by the SH0ES collaboration using the Cepheid-calibrated cosmic distance ladder, from which measurement yields $H_0 = (73.2\pm 1.3) \, km/(s. Mpc)$~\cite{Riess:2020fzl}.
In fact, there exist a {\em plethora} of different techniques not affected by Planck data for calibrating $\Lambda$CDM, through which the value of $H_0$ can also be inferred.
The so-called {\em early-Universe calibrations} combining, for instance, measurements of Big Bang nucleosynthesis (BBN) with data from baryonic acoustic oscillations (BAO) \cite{2019JCAP...10..029S,2013MNRAS.436.1674A,2015PhRvD..92l3516A,Blomqvist:2019rah,2019JCAP...10..044C} or with supernovae constraints \cite{2017MNRAS.467..731V,2021PhRvD.103j3533B}, all lead to $H_0$ values below $70 \, km/(s. Mpc)$.
Such value is confronted with another {\em plethora} of direct measurements of the Universe's local expansion rate which removes any bias eventually introduced from Cepheid observations. 

As discussed in Refs.~\cite{Lesgougues,Freedman2019jwv,Huang2019yhh,Khetan2020hmh,Hubble1,Hubble2,Hubble3,4444,5555,6666,7777,8888,9999,pheno}, the data analysis of the vast majority of such experiments yields $H_0$ values significantly closer to the value obtained by SH0ES, being followed by accomplished experimental efforts which suggest values of $H_0$ systematically larger than the value inferred by Planck and do not exclude explanations for the conflicting $H_0$ values supported by yet unknown systematic effects \cite{Hubble1,Hubble2,Hubble3}.
Considering that the systematic effect analysis seems to not be sufficient for fixing such a Hubble tension issue, corrections to the $\Lambda$CDM model operating either in the {\em early} or in the {\em late} time Universe have been considered.

Of course, modifications on the physics of the early universe are not so welcome, either by theoretical or by phenomenological reasons. 
In this letter, the explanation for Hubble tension is discussed in the framework of phase-space quantum mechanics extended to (quantum) cosmology. This would impact the background physics of the early Universe without impacting its phenomenological (classical) outputs interpreted as effectively measured smeared-out quantities.
Reporting about the nature of the problem, a generalized formulation constructed upon the Weyl-Wigner phase-space framework is shown to explain and eliminate the divergence between early- and late- time predictions for $H_0$.
Exclusively considering one-parameter dependent corrections from a quantum origin, which are mediated by generalized (localized) phase-space quantum states, predictions for $H_0$ are shown to smoothly transit from early- and late- time phenomenologically obtained values, hence resolving the Hubble tension issues.

The outline of the letter is as it follows. A summary of the Hubble tension from a theoretical perspective is presented in Section II. The phase-space quantum mechanics framework and the corresponding modifications on the Friedmann equation and on the evaluation of the Hubble parameter $H(z)$ in terms of an effective quantum potential is discussed in Section III.
Quantum corrections and the Hubble tension solution are obtained in Section IV. Finally, our conclusions are drawn in Section V.

\section{Summary of the Hubble tension from the theoretical perspective}

The observed angular scale of sound horizon at recombination, $\theta_s =(1.04109\pm 0.00030)\times 10^{-2}$, is the most precisely obtained parameter from CMB measurements and, therefore, a key scale in the $\Lambda$CDM phenomenology.
The multipole moment $(l_s)$ of the first acoustic peak determines the angle subtended by the sound horizon at the last scattering (LS) surface, such that a correspondence between the angular variation $\theta$ of a spherical harmonic of multipole $l$ is set as $\ell_s \simeq 2/\theta_s$.
The angle subtended by the sound horizon, $\theta_s$, is related to two cosmological distance definitions: the angular-diameter distance, $D_A$, and the comoving sound horizon, $r_s$, by $\theta_s= r_s/D_A$.

Considering the sound speed of the baryon-photon fluid, $c_s = dp/d\rho$, where $\rho$ and $p$ are energy density and pressure, respectively, the comoving sound horizon, $r_s$, is obtained by integrating the sound speed of the photon-baryon fluid, $c_s(z)$, about the redshift $z$ as 
\begin{equation}\label{eq:soundhorizon}
    r_s = \int_{z_{_{\rm LS}}}^\infty \frac{c_s(z)\, dz}{H(z)} = \frac{c}{\sqrt{3}H_{_{\rm LS}}} \int_{z_{_{\rm LS}}}^\infty {dz} {\left[\frac{\rho(z)}{\rho(z_{_{\rm LS}})}  \left(1 + \frac{3\omega_b}{4\omega_\gamma}\frac{1}{1+z}\right) \right]^{-1/2}},
\end{equation}
where
\begin{equation}
 c_s(z) = c \left[ 3\left(1+\frac{3\omega_b}{4\omega_\gamma}\frac{1}{1+z}\right) \right]^{-1/2}, \label{sound}    
\end{equation}
$\omega_b= \Omega_b h^2$ is the physical baryon density today, i.e. $\omega_b=0.0224\pm0.0001$, determined by the higher-peak structure in the CMB power spectrum, and $\omega_\gamma =2.47 \times 10^{-5}$ is the physical photon energy density from Planck's $\Lambda$CDM \cite{4444}.
The expansion rate at the CMB photon last scattering redshift ($z_{_{\rm LS}}\simeq 1080$) is
\begin{equation}
    H_{_{\rm LS}} = H_0 \,h^{-1} (1+z_{_{\rm LS}})^2 \sqrt{ \omega_r +\frac{\omega_m}{1+z_{_{\rm LS}}} },
\label{HLS}    
\end{equation}
where $h\equiv H_0/(100\, {\rm km}\, {\rm sec}^{-1}\, {\rm Mpc}^{-1})$ is the dimensionless form for the Hubble constant, $\omega_m=\Omega_m h^2$ is the physical nonrelativistic-matter density today, $\omega_m = 0.142\pm 0.001$ (also fixed fairly precisely by the higher-peak structure in the CMB),
and $\omega_r$ is the physical radiation density,
\begin{equation}\label{neu}
     \omega_r = \omega_\gamma\left[ 1 + \frac78 N_{\rm eff} \left( \frac{4}{11} \right)^{4/3} \right],
\end{equation}
where the second term accounts for three neutrino mass eigenstates ($N_{\rm eff}\sim3.06$ \cite{Dodelson}).

The complementary cosmological timeline is depicted by the angular-diameter distance
\begin{equation}
    D_A = \frac{c}{H_0} \int_{0}^{z_{_{\rm LS}}}
    \frac{dz}{ \left[\rho(z)/\rho_0 \right]^{1/2}},
\label{angulardiameterdistance}
\end{equation}
the (comoving) angular-diameter distance to the LS surface.
For Eq.~\eqref{eq:soundhorizon}, the cosmic inventory contributions are resumed by the early-Universe energy density written as $\rho(z) \propto \omega_m (1+z)^3 + \omega_r(1+z)^4$, for which tiny dark energy contributions ($z > z_{_{\rm LS}}$) are ignored.
For Eq.~\eqref{angulardiameterdistance}, however, the total energy density, $\rho(z)$, is now relevant for the period
from recombination to the present time, $z=0$. In this case, $\rho(z)/\rho_0 \sim \Omega_m(1+z)^3+(1-\Omega_m)(1+z)^{-3(1+w)}$, where a dark-energy equation-of-state parameter, $w=-1$, can be introduced.

Finally, from $\theta_s = r_s/D_A$, one infers the Hubble constant at present, $H_0$, from
\begin{eqnarray}
     H_0 &=& \sqrt{3} H_{_{\rm LS}} \theta_s \frac{  \int_{0}^{z_{_{\rm LS}}}
    dz\, \left[\rho(z)/\rho_0 \right]^{-1/2}   }
     {  \int_{z_{_{\rm LS}}}^\infty dz\, \left[\rho(z)/\rho(z_{_{\rm LS}}) \right]^{-1/2} \left[1 + {3\omega_b}/({4\omega_\gamma}{(1+z)})\right]^{-1/2}     },
 \label{cmbH0}
\end{eqnarray}
from which an implicit form for determining $h=H_0/(100 {\rm km}~{\rm sec}^{-1}~{\rm Mpc}^{-1})$ is written as a constraint equation,
\small{\begin{eqnarray}
     1  &=&\sqrt{3}  (1+z_{_{\rm LS}})^2 \sqrt{ \omega_r +\frac{\omega_m}{1+z_{_{\rm LS}}} } \theta_s \frac{  \int_{0}^{z_{_{\rm LS}}}
    dz\, \left[\omega_m(1+z)^3+(h^2-\omega_m)(1+z)^{-3(1+w)} \right]^{-1/2}}
     { \int_{z_{_{\rm LS}}}^\infty {dz} { \left[\rho(z)/\rho(z_{_{\rm LS}}) \right]^{-1/2}  \left[1 + {3\omega_b}/({4\omega_\gamma}{(1+z)})\right]^{-1/2}} },
 \label{cmbhfake}
\end{eqnarray}}\normalsize
if put in terms of $\omega_{b,m,r,\gamma}$.
In this approach \cite{Lesgougues}, Eq.~\eqref{cmbH0} has been used, $\rho_\mathrm{Crit} \equiv \rho_{0}$, and finding $h$ is reduced to the calibration of the coefficient of late-Universe dark energy contribution.
This summarizes the procedures for crudely computing $h(H_0)\equiv h_{_{LT}}$ and, of course, eventual discrepancies between the results for different cosmological times.

\section{Phase-space quantum mechanics and the effective quantum potential}

Phase-space quantum mechanics described in terms of the WW framework \cite{Wigner,Ballentine,Case} encompasses all the features of a quantum system through a {\em quasi}-probability distribution: the so-called Wigner function, $\mathcal{W}(x,\, k)$. The Wigner function is given in terms of canonical coordinates of position, $x$, and momentum, $k$, through the Weyl transform of the quantum density matrix operator, $\hat{\rho} = |\psi \rangle \langle \psi |$, written as
\begin{equation}
\hat{\rho} \to \mathcal{W}(x,\, k) = \pi^{-1} 
\int^{+\infty}_{-\infty} \hspace{-.35cm}dq\,\exp{\left[2\, i \, k \,q\right]}\,
\psi(x - q)\,\psi^{\ast}(x + q),\label{222}
\end{equation}
in this case, cast in a dimensionless form, such that the reduced Planck constant, $\hbar$, has been set equal to $1$.

The probability flux is driven by the continuity equation\cite{Case,Ballentine,Steuernagel3,NossoPaper,Meu2018},
\begin{equation}\label{z51dim}
{\partial_{\tau} \mathcal{W}} + {\partial_x \mathcal{J}_x}+{\partial_k \mathcal{J}_k} = {\partial_{\tau} \mathcal{W}} + \mbox{\boldmath $\nabla$}_{\xi}\cdot\mbox{\boldmath $\mathcal{J}$} =0,
\end{equation}
where the time variable, $\tau$, is also dimensionless. For Hamiltonians cast as $\mathcal{H}(x,\,k) = \mathcal{K}(k) + \mathcal{V}(x)$,   
the corresponding Wigner currents are then given by \cite{Novo2021A}
\begin{eqnarray}
\label{imWAmm}\mathcal{J}_x(x, \, k;\,\tau) &=& +\sum_{\eta=0}^{\infty} \left(\frac{i}{2}\right)^{2\eta}\frac{1}{(2\eta+1)!} \, \left[\partial_k^{2\eta+1}\mathcal{K}(k)\right]\,\partial_x^{2\eta}\mathcal{W}(x, \, k;\,\tau),\\
\label{imWBmm}\mathcal{J}_k(x, \, k;\,\tau) &=& -\sum_{\eta=0}^{\infty} \left(\frac{i}{2}\right)^{2\eta}\frac{1}{(2\eta+1)!} \, \left[\partial_x^{2\eta+1}\mathcal{V}(x)\right]\,\partial_k^{2\eta}\mathcal{W}(x, \, k;\,\tau),
\end{eqnarray}
from which the contributions from $\eta \geq 1$ in the corresponding series expansions depict the quantum back reaction.

Once truncated at $\eta = 0$, currents from Eqs.~\eqref{imWAmm} and \eqref{imWBmm} reproduce to the classical Liouvillian regime \cite{Case,Ballentine}, with the associated phase-space vector velocity identified by $\mathbf{v}_{\xi(\mathcal{C})} = \dot{\mbox{\boldmath $\xi$}} = (\dot{x},\,\dot{k})\equiv ({\partial_k \mathcal{H}},\,-{\partial_x \mathcal{H}})$. 
Likewise, an effective quantum regime can be connected to a parametric definition of a quantum-analog velocity, $\mathbf{w}$, implicitly given in terms of the vector currents, $\mbox{\boldmath $\mathcal{J}$} = \mathbf{w}\,\mathcal{W}$, eith $\mbox{\boldmath $\nabla$}_{\xi}\cdot\mbox{\boldmath $\mathcal{J}$} = \mathcal{W}\,\mbox{\boldmath $\nabla$}_{\xi}\cdot\mathbf{w}+ \mathbf{w}\cdot \mbox{\boldmath $\nabla$}_{\xi}\mathcal{W}$
{\color{purple}\footnote{\color{purple} Stationary and Liouvillian patterns are resumed by flow divergent properties,
\begin{equation} \label{helps}
\mbox{\boldmath $\nabla$}_{\xi}\cdot\mbox{\boldmath $\mathcal{J}$} = \sum_{\eta=0}^{\infty}\frac{(-1)^{\eta}}{2^{2\eta}(2\eta+1)!} \, \left\{
\left[\partial_x^{2\eta+1}\mathcal{V}(x)\right]\,\partial_k^{2\eta+1}\mathcal{W}
-
\left[\partial_k^{2\eta+1}\mathcal{K}(k)\right]\,\partial_x^{2\eta+1}\mathcal{W}
\right\},\end{equation}
and
\begin{equation}\label{div2}
\mbox{\boldmath $\nabla$}_{\xi} \cdot \mathbf{w} = \sum_{\eta=0}^{\infty}\frac{(-1)^{\eta}}{2^{2\eta}(2\eta+1)!}
\left\{
\left[\partial_k^{2\eta+1}\mathcal{K}(k)\right]\,
\partial_x\left[\frac{1}{\mathcal{W}}\partial_x^{2\eta}\mathcal{W}\right]
-
\left[\partial_x^{2\eta+1}\mathcal{V}(x)\right]\,
\partial_k\left[\frac{1}{\mathcal{W}}\partial_k^{2\eta}\mathcal{W}\right]
\right\}, ~~~\end{equation}
respectively.}}. Stationary and classical global patterns, with $\mbox{\boldmath $\nabla$}_{\xi}\cdot\mbox{\boldmath $\mathcal{J}$}=0$ and $\mbox{\boldmath $\nabla$}_{\xi} \cdot \mathbf{w}=0$, respectively \cite{Meu2018}, come from $\eta=0$ contributions, which is consistent with the limit of $\mathbf{w} \to \mathbf{v}_{\xi(\mathcal{C})}$.  

From such results, an effective quantum modification to the classical potential, $\mathcal{V}(x) \to \mathcal{U}(x)$, which encompasses all the non-linear effects introduced by $\partial_x^{2\eta+1}\mathcal{V}(x)$ contributions in the series expansion from Eq.~\eqref{imWBmm}, can be obtained by replacing 
${v}_{k(\mathcal{C})} = \dot{k}\equiv -{\partial_x \mathcal{H}}=-{\partial_x \mathcal{V}}$ by
\begin{eqnarray}
\label{imWBmmw} w_k(x, \, k;\,\tau) = -\partial_x \mathcal{U}&=& -\sum_{\eta=0}^{\infty} \left(\frac{i}{2}\right)^{2\eta}\frac{1}{(2\eta+1)!} \, \left[\partial_x^{2\eta+1}\mathcal{V}(x)\right]\,\frac{\partial_k^{2\eta}\mathcal{W}(x, \, k;\,\tau)}{\mathcal{W}(x, \, k;\,\tau)}.
\end{eqnarray}
Of course, the above proposal, Eq.~\eqref{imWBmmw}, works fine only for $\partial_k^{2\eta}\mathcal{W}(x, \, k;\,\tau)/\mathcal{W}(x, \, k;\,\tau)$ matching the conditions which leads to $\mathcal{U} \equiv \mathcal{U}(x)$.
This can only be achieved through convergent series criteria applied to very particular sets of Wigner functions, $\mathcal{W}(x, \, k;\,\tau)$, which fits adequate phenomenological patterns. This shall be discussed in the following.

\section{Quantum corrections for sinusoidal Wigner distributions}

By replacing the Wigner distribution at Eq.~\eqref{imWBmmw} by a sinusoidal function of the product $k\,x$ modulated by an $x$ and $\tau$-dependent arbitrary function, $g(x;\,\tau)$, such that $\mathcal{W}(x, \, k;\,\tau) = g(x;\,\tau)\, \mbox{Sn}(\mu \,k\,x)$, where $\mu$ is an arbitrary constant parameter, and $\mbox{Sn}(\dots)$ is identified as either $\sin(\dots)$ or $\cos(\dots)$, after some straightforward math manipulations, one obtains
\begin{eqnarray}\label{imWBmmws1}
\partial_x\mathcal{U}(x)&=& \sum_{\eta=0}^{\infty} \left(\frac{\mu\,x}{2}\right)^{2\eta}\frac{1}{(2\eta+1)!} \, \left[\partial_x^{2\eta+1}\mathcal{V}(x)\right].
\end{eqnarray}
For the classical potential $\mathcal{V}(x)$ written as a sum of polynomial and inverse polynomial contributions, $\mathcal{V}(x) = \sum_{\kappa} a_\kappa x^{-\kappa}$, with $\kappa \in \mathbb{Z}$, Eq.~\eqref{imWBmmws1} can be cast as
\begin{eqnarray}\label{imWBmmws2}
\partial_x\mathcal{U}(x)&=& - \sum_{\kappa} \left\{a_\kappa \sum_{\eta=0}^{\infty} \left(\frac{\mu}{2}\right)^{2\eta}\frac{(2\eta+\kappa)!}{(\kappa-1)!(2\eta+1)!}\right\} \,x^{-(\kappa+2\eta+1)+2\eta}\nonumber\\
&=& - \sum_{\kappa} \left\{a_\kappa \sum_{\eta=0}^{\infty} \left(\frac{\mu}{2}\right)^{2\eta}\frac{\Gamma(2\eta+\kappa+1)}{\Gamma(\kappa)\Gamma(2\eta+2)}\right\} \,x^{-(\kappa+1)},
\end{eqnarray}
in order to evince additional simplifications that, after a straightforward integration in $x$, result in
\begin{eqnarray}\label{imWBmmws3}
\mathcal{U}(x)&=& \sum_{\kappa} \left\{a_\kappa \sum_{\eta=0}^{\infty} \left(\frac{\mu}{2}\right)^{2\eta}\frac{\Gamma(2\eta+\kappa+1)}{\Gamma(\kappa+1)\Gamma(2\eta+2)}\right\} \,x^{-\kappa}.
\end{eqnarray}
Eq.~\eqref{imWBmmws3} can then be cast in the form of
\begin{eqnarray}\label{imWBmmws4}
\mathcal{U}(x)&=& \sum_{\kappa} b_\kappa x^{-\kappa},
\end{eqnarray}
with
\begin{eqnarray}\label{imWBmmws5}
b_\kappa &=& a_\kappa \sum_{\eta=0}^{\infty} \left(\frac{\mu}{2}\right)^{2\eta}\frac{\Gamma(2\eta+\kappa+1)}{\Gamma(\kappa+1)\Gamma(2\eta+2)}\nonumber\\
&=&\frac{2^\kappa a_\kappa}{\mu\,\kappa}\left[(2-\mu)^{-\kappa}-(2+\mu)^{-\kappa}\right], \end{eqnarray}
which constrains $\mu$ to $\mu \in(-2,\,+2)$ (with $\kappa \in \mathbb{R}$) in order to not introduce relevant topological changes to the effective quantum potential as, of course, it should be expected from the convergence criterium for the above series expansion, which, in particular, exhibits analytical continuation for $\mu = 0$.

To summarize, three distinguished features must be identified from the above manipulations:

$i)$ $x$-modulated sinusoidal (wave and Wigner) functions describe large and variable classes of quantum phenomena.
Proposing test Wigner functions like sinusoidal ones means that one is probing the quantum scenario without specific constraints of a particular quantum model.

$ii)$ For classical potentials written $\mathcal{V}(x) = \sum_{\kappa} a_\kappa x^{-\kappa}$, which naturally also encompass an enormous set of classical and quantum integrable systems, the quantum modifications are evidently identified as a reconfiguration of the polynomial (and inverse polynomial) contributions identified by the modified coefficients $b_\kappa$ at $\mathcal{U}(x) = \sum_{\kappa} b_\kappa x^{-\kappa}$ (cf. Eq.~\eqref{imWBmmws4}).
Such property leads to a multiple re-scaling of the $x$ coordinate driven by each $\kappa$ index contributions from the classical potential, which, as it shall be quantified in the following subsections, seems to be essential in producing modification patterns as those expected for the Hubble tension problem.

$iii)$ To construct Wigner (and corresponding position localized wave) functions as generic as one would like, normalized configurations demand for finite and even parity momentum configurations at the phase-space, i.e. $\mathcal{W}(x, \, k;\,\tau) = g(x;\,\tau)\, \mbox{Sn}(\mu \,k\,x)$ should be more properly written as $\mathcal{W}(x, \, k;\,\tau) = \theta(k - k_0)\theta(k + k_0)\,g(x;\,\tau)  \mbox{cos}(\mu \,k\,x)$ for integration manipulations such that the associated coordinate probability distribution would be given by
\begin{eqnarray}
\vert\psi(x;\,\tau)\vert^{2}&=& 
\int^{+\infty}_{-\infty} \hspace{-.35cm}dk\,\mathcal{W}(x, \, k;\,\tau)\nonumber\\
&=& 
\int^{+\infty}_{-\infty} \hspace{-.35cm}dk\,\theta(k - k_0)\theta(k + k_0)\,g(x;\,\tau)\,\mbox{cos}(\mu \,k\,x)\nonumber\\
&=& 
2{g(x;\,\tau)}\,\frac{\mbox{sin}(\mu\, k_0\,x)}{\mu\, x},\label{2121}
\end{eqnarray}
which is finite and continuous at $x=0$, with $\psi(x;\,\tau)$, normalization guaranteed by an even function in $x$, $g(x;\,\tau)$. In the momentum space, the mathematical manipulations are similar to those ones for the coordinate space in the discussion of Wigner phase-space solutions of the infinite squared well potential problem \cite{Belloni}. 
In particular, the finite momentum scenario has also been considered in the description of quark-gluon plasma \cite{Fili19}, in kinetic theories for massive and massless fermions \cite{Xin20}, and more generically, in the investigation of (in)finite quantum system with a discrete and (in)finite-dimensional Hilbert space \cite{Inf03,Inf02,Inf01,Inf00}.

\section{Quantum corrections and the Hubble tension solution}

Departing from the Einstein-Hilbert action given by
\begin{equation}
\mathcal{S} = \frac{1}{16 \pi G}\int{d^{4}x\, \sqrt{-\mathtt{g}}\, \mathcal{R}},
\label{eqn00}
\end{equation}
the scalar curvature is identified by $\mathtt{g} = \det{(g_{\mu \nu})}$, $\mathcal{R} = R^{\mu \nu}\,g_{\mu \nu}$, $\sqrt{-\mathtt{g}} = N(t) \, a(t)^3$, where the lapse function, $N(t)$, and the scale parameter, $a(t)= (1+z(t))^{-1}$, are arbitrary non-vanishing functions of time, $t$, and the line element of the Robertson-Walker (RW) metric in an homogeneous and isotropic space-time where
\begin{equation}
ds^{2} = - \sigma^{2}\left[N(t)^{2}\,dt^{2} - a(t)^2\,\left(\frac{dr^2}{1- {\Omega_{_{C}}} r^2} + r^2\,d\vartheta^{2}\right) \right],
\label{eqn01}
\end{equation}
is the most general form of a $SO(4)$-invariant metric in a $M = \mathbb{R} \times S^3$ topology \cite{BertolamiMourao}.
The above notation considers $c = \hbar = 1$, $\sigma$ as a normalization constant, and ${\Omega_{_{C}}} =0,\, +1$, and $-1$ denoting the curvature corresponding to $\mathbb{R}^3$, $S^3$ and $H^3$ hypersurfaces, all for $d\vartheta^{2} = d\theta^2 + \sin{(\theta)}^2 d\phi^2$.

In terms of metric components, the above quantities are all obtained from three-dimensional quantities used to describe the general relativity (GR) in the Arnowitt-Deser-Misner (ADM) formalism \cite{Misner,ABOB01}:
$g_{ij}$, $\Pi_{ij} = \sqrt{-g}\left(\Gamma_{kl}^0 - g_{kl}\,\Gamma_{mn}^0\, g^{mn}\right)g^{ik}g^{jl}$,
$N = (-g^{00})^{-{1}/{2}}$, and the shift vector, $N_{i} = g_{0i}$, with the connection, $\Gamma_{ij}^{k}$, as an independent quantity, and with {\em latin} indices running from $1$ to $3${\color{purple}\footnote{\color{purple} For completeness, the extrinsic curvature is written as
\begin{equation}
K_{ij}=\frac{1}{2\sigma N}\left(-\frac{\partial g_{ij}}{\partial t}+\nabla_{i}N_{j}+\nabla_{j}N_{i}\right),
\label{eqn03}
\end{equation}
with $\nabla_{i}$ denoting the $3$-dimensional covariant derivative, for the metric Eq.~\eqref{eqn01},
\begin{equation} 
K_{ij}=-\frac{1}{\sigma N} \frac{\dot{a}}{a} g_{ij}, \quad \mbox{with} \quad K = K^{ij}g_{ij} = -\frac{3}{\sigma N} \frac{\dot{a}}{a},
\label{eqn04}
\end{equation}
in case of $N_i = 0$ (and $\Omega_{_{C}}= 1$).}}.
For $N_i = 0$ and $\Omega_{_{C}}= 1$, the $3$-dim Ricci tensor components and the corresponding Ricci scalar are given, respectively, by $R_{ij}={2 g_{ij}}/{(\sigma^{2}a^{2})}$ and $R={6}/{\sigma^{2}a^{2}}$, which can be assumed as the cornerstone of the simplified discussion of the quantum mechanical problem according to the Wheeler-DeWitt (WDW) framework \cite{DeWitt67,Hartle83,Linde84}. In this case, a minisuperspace action can be cast in the form of
\begin{equation}
S_{SM}=\frac{1}{2}\int dt \left(\frac{N}{a}\right)\left[-\left(\frac{a}{N}\dot{a}\right)^{2} +{\Omega_{_{C}}}a^{2} -\Omega_{_{\Lambda}}a^{4}-\Omega_{_{r}}-{\Omega_{_{m}}}{a} \right],
\label{eqn10SM}
\end{equation}
from which one notices that ${\Omega_{_{C}}} > 0$ stands for the curvature coupling constant and the sign of $\Omega_{_{\Lambda}}$ follows the sign of the cosmological constant. As before, $\Omega_{_{m}}$ and $\Omega_{_{r}}$, are associated to matter and radiation contributions, respectively.
From $S_{SM}$, the minisuperspace Hamiltonian density can then be written as
\begin{equation}
 H=\Pi_{a}\dot{a}-\mathcal{L}=\frac{1}{2}\frac{N}{a}\left(-\Pi_{a}^{2}-{\Omega_{_{C}}}a^{2} +\Omega_{_{\Lambda}}a^{4}+\Omega_{_{r}}+{\Omega_{_{m}}}{a^{2}}\right),
\label{eqn12SM}
\end{equation}
where the canonical conjugate momentum associated to $a$ is identified by
$\Pi_{a}={\partial \mathcal{L}}/{\partial \dot{a}} = - {a}\dot{a}/{N}$.
Following the canonical quantization procedure \cite{DeWitt67,Hartle83}, the momentum $\Pi_{a}$ is read as an operator \cite{Hartle83}, $\Pi_{a}\mapsto -i({d}/{d a})${\color{purple}\footnote{\color{purple} Such that $$\quad\Pi_{a}^{2}=-\frac{1}{a^{q}}\frac{d}{d a}\left(a^{q}\frac{d}{d a}\right),$$ where the choice of $q$ does not affect the semiclassical analysis \cite{KolbTurner:1989}.}}, as to have
the minisuperspace Hamiltonian operator acting as
\begin{equation}\label{eqn14}
\frac{1}{2}\left(\frac{d^{2}}{d a^{2}}-{\Omega_{_{C}}}a^{2} +\Omega_{_{\Lambda}}a^{4}+\Omega_{_{r}}+{\Omega_{_{m}}}{a}\right)\tilde{\psi}(a)=0,
\end{equation}
i.e. the WDW equation for the wave function of the Universe, $\tilde{\psi}(a)$.
An effective potential is then identified by 
\begin{equation}
V(a)=\frac{1}{2}\left({\Omega_{_{C}}}a^{2}-\Omega_{_{\Lambda}}a^{4}-\Omega_{_{r}}-{\Omega_{_{m}}}{a}\right),
\label{eqn15SM}
\end{equation}
which sets the ingredients for the quantum Wigner framework analysis, where the classical potential, $\mathcal{V}(x)$, is related to $V(a)$ by $ \mathcal{V}(x) = - x^{-\sigma} V^{({\Omega_{_{C}}}\to 0)}(a)$ (with $x \equiv a$), being identified by
\begin{eqnarray}
\mathcal{V}(x) = x^{(4-\sigma)}\rho(x)/\rho_\mathrm{Crit} &=& x^{-\sigma}\left(\Omega_{_{\Lambda}}x^{4}+\Omega_{_{r}}+{\Omega_{_{m}}}{x}\right)\nonumber\\
&=&x^{-\sigma}h^{-2}\left((h^2-\omega_{_{r}}-\omega_{_{m}})\, x^{4}+\omega_{_{r}}\,+{\omega_{_{m}}}\,{x}\right),
\label{eqn17SM}
\end{eqnarray}
with $\sigma$ constrained by the choice of the lapse function, $N = - a^{1-\sigma}$.
Notice that the curvature contribution has being ignored from this point since it is not relevant to such a preliminar analysis. 
From Eq.~\eqref{imWBmmws4}, one would have
\begin{eqnarray}\label{imWBmmws4QC}
\mathcal{U}(x)&=& x^{-\sigma}\left(\Omega_{_{\Lambda}}\,b_{\sigma-4}(\mu)\, x^{4}+\Omega_{_{r}}\,b_{\sigma}(\mu)\,+{\Omega_{_{m}}}\,b_{\sigma-1}(\mu)\,{x}\right)\nonumber\\
&=& x^{-\sigma}h^{-2}\left((h^2-\omega_{_{r}}-\omega_{_{m}})\,b_{\sigma-4}(\mu)\, x^{4}+\omega_{_{r}}\,b_{\sigma}(\mu)\,+{\omega_{_{m}}}\,b_{\sigma-1}(\mu)\,{x}\right),
\end{eqnarray}
such that the coefficients $b_\kappa(\mu)$, given by Eq.~\eqref{imWBmmws5}, are one parameter functions of $\mu$, which modulate the modifications on the cosmic energy density pattern.

Since the quantum effects are assumed to be suppressed at very late times ($x \lesssim 1$) in order to not drastically change the critical density, $\rho_\mathrm{Crit}$, the coefficient $b_{\sigma-4}(\mu)$ is set equal to unity. Expecting fixed output values at $z=0$, $\tilde{h}(a_0=1) = h_{_{LT}} = 0.732$, one could read the results from quantum corrections, Eq.~\eqref{imWBmmws4QC}, as $\mathcal{U}(x)= (\tilde{h}/h_{_{LT}})^{-2} \mathcal{V}(x)$ ($\equiv \tilde{\rho}(x)= (\tilde{h}/h_{_{LT}})^{-2}\rho(x)$), which would be (mis)interpreted as a modulation from the Hubble parameter, $\tilde{h}$, at early times.

\begin{figure}[h!]
\includegraphics[scale=.24]{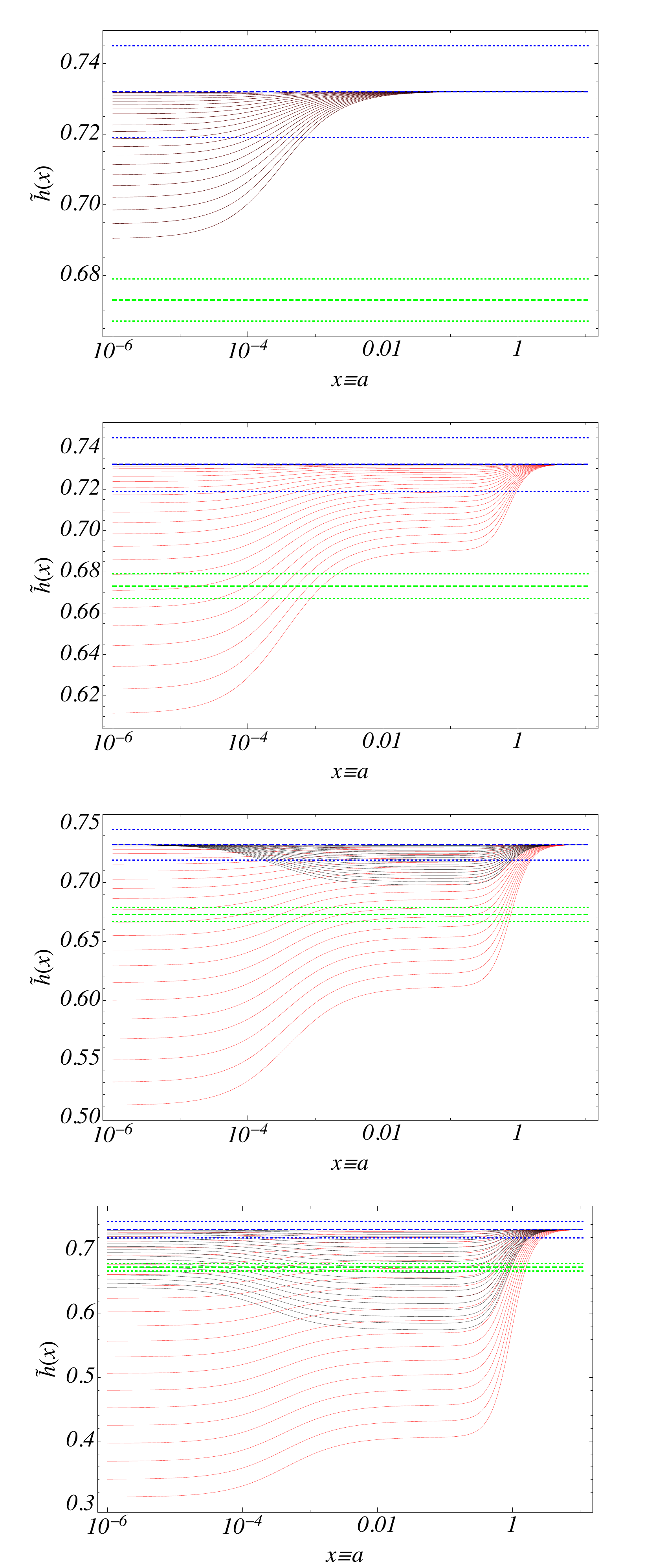}
\renewcommand{\baselinestretch}{1}
\caption{\footnotesize (Color online) Results for $ \tilde{h}(x) = h_{_{LT}} (\mathcal{V}(x)/\mathcal{U}(x))^{1/2}$ as function of the scale factor, $x$, for $\mu$ varying from $0.1$ to $1.1$ (with steps of 0.05, for plot lines from top to bottom). Plots are for integer values of $\sigma$, with $\vert\sigma\vert = 0$, $1$, $2$ and $4$ (from top to bottom). Red lines are for $\sigma \geq 0$ and black lines are for $\sigma \leq 0$. Blue (green) lines are for $\tilde{h}= 0.732 \pm 0.013\, (0.673 \pm 0.006)$.}
\label{primeira}
\end{figure}

\begin{figure}[h!]
\includegraphics[scale=.45]{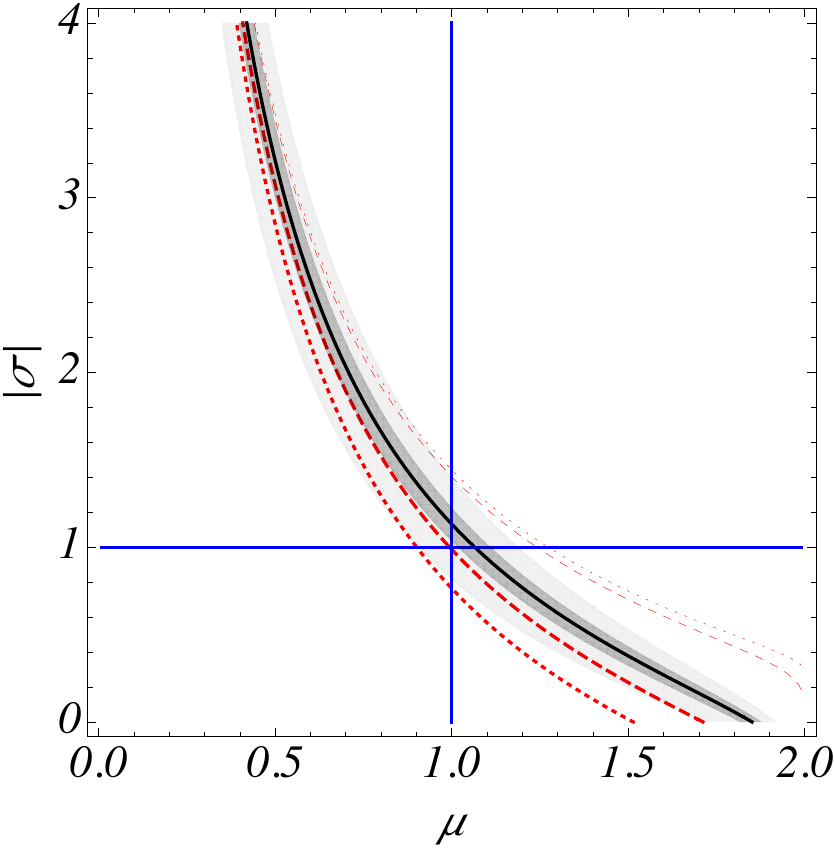}
\includegraphics[scale=.333]{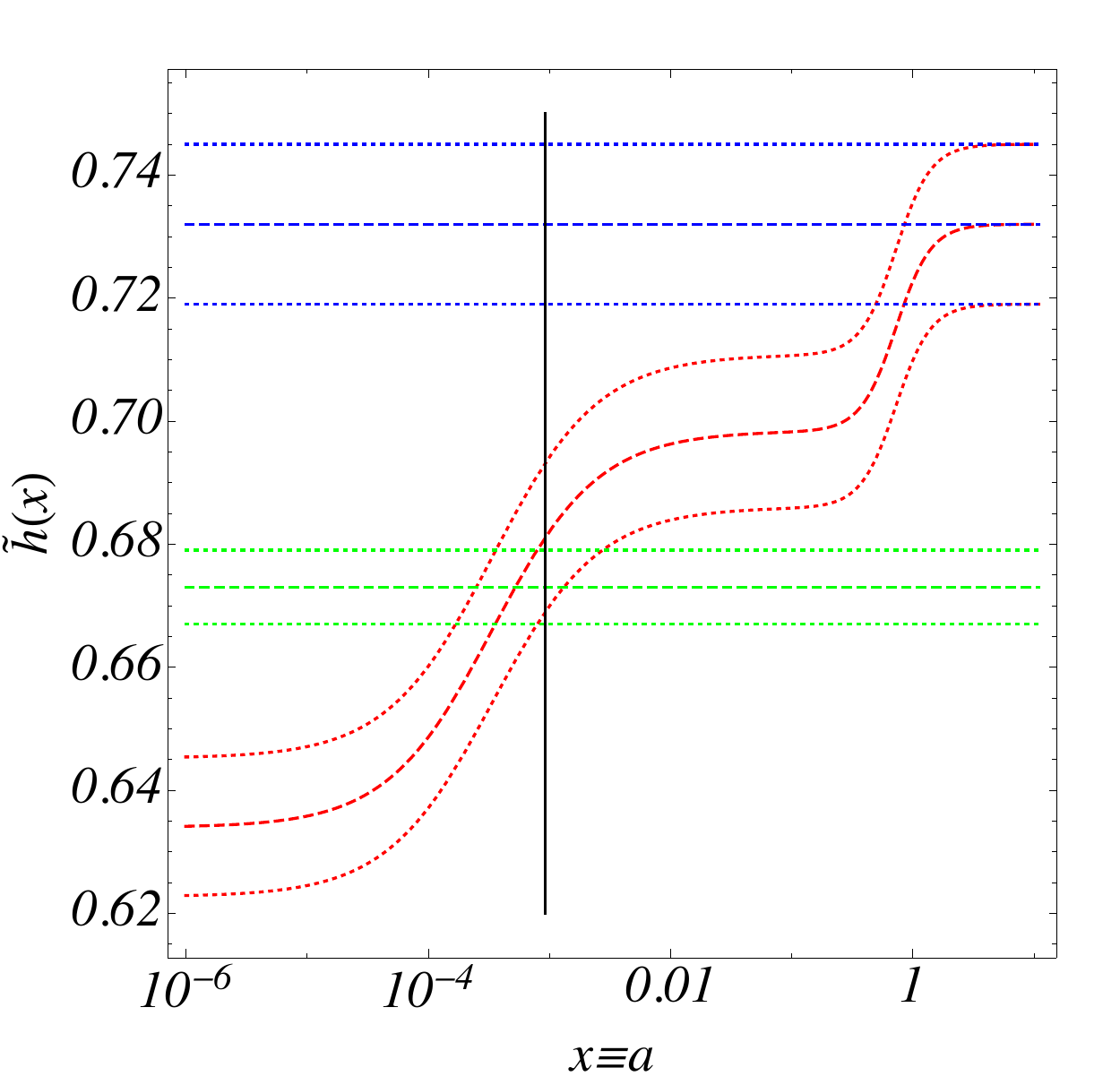}
\renewcommand{\baselinestretch}{1}
\caption{\footnotesize
(Color online) (First Plot) $\sigma$ and $\mu$ constraint such that $\tilde{h}(x\sim1) \to h = 0.732$ and $\tilde{h}(10^{-4} \lesssim x \lesssim 10^{-1}) \sim 0.673$ (black thick line). Dark gray region shows the limits when experimental errors for early time measurements, i.e. $\tilde{h}= 0.673 \pm 0.060$, are included. Light gray region shows the limits when experimental errors for both early and late time measurements, $\tilde{h} = 0.673 \pm 0.006$ and $\tilde{h}\to h_{_{LT}} = 0.732 \pm 0.013$, are included. Red dashed and dotted lines depict the smooth changes of $\tilde{h}(x)$ as function of $x$, in correspondence with results from Fig.~\ref{primeira}. In this case, $\sigma$ and $\mu$ constraints are for $\tilde{h}(10^{-1})$ (dotted thin), $\tilde{h}(10^{-2})$ (dashed thin), $\tilde{h}(5 \times 10^{-4})$ (dashed thick) and $\tilde{h}(2 \times 10^{-4})$(dotted thick).  (Second Plot) Results for $\tilde{h}(x) = h (\mathcal{V}(x)/\mathcal{U}(x))^{1/2}$ as function of the scale factor, $x$, for $\mu=\sigma =1$. Blue (green) lines are for $\tilde{h}= 0.732 \pm 0.013\, (0.673 \pm 0.006)$. Black line mark $z = 1080$.}\label{terca}
\end{figure}

Fig~\ref{primeira} depicts the results for $\tilde{h}^2 \to  \tilde{h}^2(x) = h_{_{LT}}^2 (\mathcal{V}(x)/\mathcal{U}(x))$ for several choices of the lapse function parameter, $\sigma$, and of the wave function parameter, $\mu$. 
The parameters were set as $h = 0.732$,  $\omega_m = 0.142\pm 0.001$ and $\omega_r$ as from Eq.~\eqref{neu}, with $\omega_\gamma =2.47 \times 10^{-5}$.
Phenomenologically expected well defined {\em plateaus} can then be identified for $\tilde{h} \sim 0.68$ and $\sim 0.72$ respectively at early and late times. As depicted in Fig.~\ref{terca}, parameters $\sigma$ and $\mu$ can be constrained one to each other as to return the phenomenologically consistent results for the Hubble parameter at early and late times, i.e. $\tilde{h}^2(x\lesssim 1) = 0.732 \pm 0.013$ and $\tilde{h}^2(x\ll 1) = 0.673 \pm 0.006$. 

Of course, the above analysis does not change the phenomenological outputs of the early-Universe physics. The input parameter, $h = 0.732$ just reflects the late-Universe classical outputs where, as expected, the quantum smeared-out effects are suppressed. The same smeared-out effects are instead effective at early times, with $\tilde{h}^2(x\ll 1) = 0.673 \pm 0.006$ identified.

Subtly, for simultaneous integer values of $\mu=\sigma=1$ (cf. crossing blue lines in Fig.~\ref{terca}), results are consistent with the phenomenology (i.e. $\tilde{h}^2(z \sim 0) = 0.732 \pm 0.013$ and $\tilde{h}^2(z \sim 1080) = 0.673 \pm 0.006$) as depicted in the second plot of Fig.~\ref{terca}. 

\section{Conclusions}

The above results are concerned with far open scenarios for quantum cosmology which, in this case, are driven by the phase space quantum mechanics tasks. Even if our analysis provides a consistent explanation for the Hubble tension, through a one-parameter dependent correction to the observed divergent results for the Hubble parameter at early and late times, this just bring up some elementary modifications to the standard cosmological model, in particular, when it is driven by a Hamiltonian approach.
Naturally, more enhanced analysis are admitted. The next step involves the computation of the angular-diameter distance, $D_A$, and the comoving sound horizon, $r_s$, by $\theta_s= r_s/D_A$ in view of the dynamical behavior of $\tilde{h}(z)$, which follows from Eqs.~\eqref{cmbH0} and \eqref{cmbhfake}.
However, one has here a free of data analysis perspective. The implications through the physical observables for quantum cosmology in the minisuperspace limit could be proposed and discussed in terms of more specified models for which exact solutions of the WDW equation could been considered and related to the phenomenology \cite{cosmoverse}. 
Concerning the approach based on phase space quantum cosmology as proposed here, the quantum modification on the Einstein-Friedmann equation arises naturally, fix the Hubble parameter divergence, and does not require any additional model assumption. As asserted, such corrections from quantum origin, once mediated by generalized (localized) phase-space quantum states, make predictions for $H_0$ smoothly fluctuate from early- and late- time phenomenologically obtained values, by resolving any Hubble tension issue.

\vspace{.5 cm}
{\em Acknowledgments -- The work was partially supported by Grant No. 2023/00392-8 (S\~ao Paulo Research Foundation (FAPESP)) and Grant No. 301485/2022-4 (CNPq).
The author would like to thank Prof. Orfeu Bertolami from Departamento de F\'isica e Astronomia, Faculdade de Ci\^{e}ncias da
Universidade do Porto, for his hospitality and for the discussions that contributed to this investigation.}

\end{document}